\documentclass[aip,graphicx,showpacs,12pt,onecolumn]{revtex4}%
\usepackage{amsmath}
\usepackage{amsfonts}
\usepackage{amssymb}
\usepackage{graphicx}%
\providecommand{\U}[1]{\protect\rule{.1in}{.1in}}
\draft
\begin{document}
\title{Josephson Amplifier for Qubit Readout}
\author{Baleegh Abdo}
\author{Flavius Schackert}
\altaffiliation{B. A. and F. S. contributed equally to this work.}

\author{Michael Hatridge}
\author{Chad Rigetti}
\altaffiliation{Current address: IBM T. J. Watson Research Center, Yorktown Heights, New York
10598, USA.}

\author{Michel Devoret}
\email{michel.devoret@yale.edu}
\affiliation{Department of Applied Physics, Yale University, New Haven, CT 06520, USA.}
\date{\today}

\begin{abstract}
We report on measurements of a Josephson amplifier (J-amp) suitable for
quantum-state qubit readout in the microwave domain. It consists of two
microstrip resonators which intersect at a Josephson ring modulator. A maximum
gain of about $20$ dB, a bandwidth of $9$ $%
\operatorname{MHz}%
$, and a center-frequency tunability of about $60$ $%
\operatorname{MHz}%
$ with gain in excess of $10$ dB have been attained for idler and signal of
frequencies $6.4$ $%
\operatorname{GHz}%
$ and $8.1$ $%
\operatorname{GHz}%
$, in accordance with theory. Maximum input power measurements of the J-amp
show a relatively good agreement with theoretical prediction. We discuss how
the amplifier characteristics can be improved.

\end{abstract}

\maketitle









%

\newpage

One of the major hurdles which scientists confront in the growing area of
solid state quantum information systems is the relatively high noise
temperature of commercial amplifiers in the microwave domain \cite{HEMT} based
on high electron mobility transistors (HEMTs), which have a noise temperature
above $2%
\operatorname{K}%
$. In order to overcome this pressing problem several low-noise amplification
schemes have been utilized, such as the use of rf-SETs
\cite{rfsetAmpl,rfsetAmpl2}, dc-SQUIDs
\cite{MicrostripDcSquidClarke,RfampldcSQUID,Hatridge}, nanoSQUIDs
\cite{nanoSQUIDs} and flux qubits \cite{OnChipAmpl}. In addition,
amplification schemes which employ large unshunted Josephson tunnel junctions
\cite{ParamYurkePRA,MovshovichPRL,JBA,Spietz,dcSquidPhasePreservingSpietz,Abdo}
have appeared recently very promising, especially two main device variations,
namely the degenerate parametric amplifier
\cite{CastellanosNat,Castellanos,Yamamoto} and the non-degenerate parametric
amplifier \cite{JPCnature,JPCnaturePhys}.

Both parametric amplifiers, the degenerate and the non-degenerate, can be
operated in either phase sensitive amplification mode where they amplify only
one quadrature of the microwave field without adding noise to the processed
signal, or in phase preserving amplification mode where they amplify both
quadratures of the microwave field at the expense of adding at least a noise
equivalent to a half input photon at the signal frequency
\cite{Caves,JPCnature}. However, only the non-degenerate amplifier has the
advantage of a full pump-signal separation in the phase preserving mode.

In a previous work by our group Bergeal \textit{et al.} \cite{JPCnature} have
demonstrated a proof of principle device based on Josephson junctions and
coplanar waveguide resonators which performs a full non-degenerate three-wave
mixing amplification of microwave signals. The device was named the Josephson
parametric converter (JPC) in order to emphasize its unique frequency
conversion property compared to other degenerate amplifiers. As a preamplifier
preceding the HEMT, the JPC had two important characteristics: it achieved
gains in excess of $40$ dB and operated near the quantum limit. In the present
work we report new results of a practical and well controlled Josephson
amplifier (J-amp) based on the JPC, which is suitable for qubit readout. To
that end, the main advantage of the present device is that it attains a high
gain-bandwidth product ($\sqrt{G}B$) of $0.1$ $%
\operatorname{GHz}%
$, where $G$ is the power gain of the device and $B$ is the $3$ dB bandwidth
at gain $G$. While a $20$ dB of gain is required in order to beat the noise
floor of the following amplifier, a large bandwidth of the order of $10$ $%
\operatorname{MHz}%
$ is essential in order to process input signals on a time scale less than
$100$ $%
\operatorname{ns}%
$.

\begin{figure}
[h]
\begin{center}
\includegraphics[
height=2.3938in,
width=5.0427in
]%
{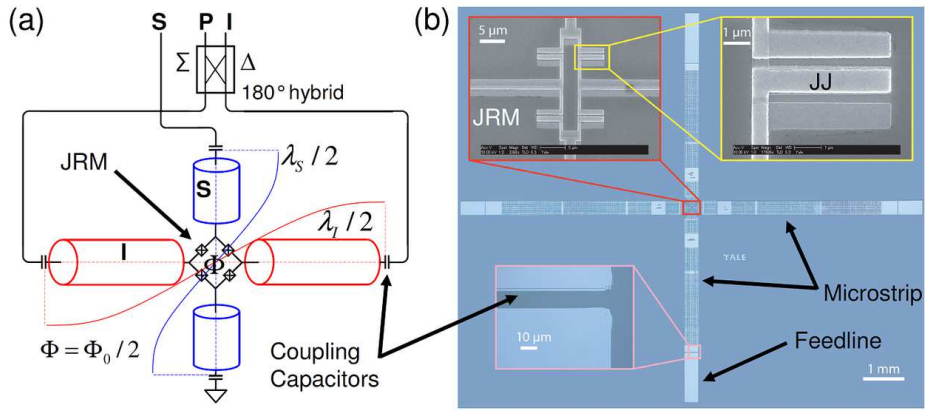}%
\caption{(Color online). (a) A circuit representation of the J-amp device
based on the JPC consisting of two half-wave transmission line resonators
($\lambda_{S}/2$ and $\lambda_{I}/2$) supporting two eigenmodes of the JRM,
denoted as signal (S) and idler (I). The idler is fed through the difference
port ($\Delta$) of a $180$ degree hybrid, while the third eigenmode denoted as
pump (P) is non-resonant and fed to the device via the sum port ($\Sigma$).
The flux bias applied to the ring is half a flux-quantum. (b) Optical
microscope image of the J-amp device, consisting of two microstrip resonators
and a JRM at the middle. The upper insets show zoom-in SEM micrographs of the
JRM (left inset) and one Josephson junction of the ring (right inset). The
bottom inset shows one of the gap capacitors which couple the feedlines to the
resonators and determine the total quality factor of the resonators.}%
\label{JPCdeviceN}%
\end{center}
\end{figure}

The J-amp consists of two half-wave microstrip resonators denoted as signal
(S) and idler (I) which intersect at an rf-current anti-node of the resonators
where a Josephson ring modulator (JRM) is incorporated. The latter element
consists of four Josephson junctions arranged in a Wheatstone bridge
configuration with half a flux quantum threading the ring (see Fig.
\ref{JPCdeviceN}). Both S and I are the differential eigenmodes of the JRM
system \cite{JPCnaturePhys,JRCAreview}. The two resonators supporting these
modes shown in Fig. \ref{JPCdeviceN} have a loaded resonance frequency
$f_{S}^{res}=8.11$ $%
\operatorname{GHz}%
$ and $f_{I}^{res}=6.44$ $%
\operatorname{GHz}%
$, a loaded quality factor $Q_{S}=100$ and $Q_{I}=64$ corresponding to
bandwidths $\kappa_{S}/2\pi=81$ $%
\operatorname{MHz}%
$ and $\kappa_{I}/2\pi=100$ $%
\operatorname{MHz}%
$ respectively.%

When operated as an amplifier, the device mixes two incoming low intensity
fields, i.e. the signal and the idler having frequencies $f_{S}$ and $f_{I}$,
with an intense microwave field called pump (P) having a frequency
$f_{P}=f_{S}+f_{I}$ and applied using the common mode of the JRM.

The measurements presented in this paper were carried out in a cryogen-free
dilution fridge at $30$ m$%
\operatorname{K}%
$. The experimental scheme used is similar to that employed in previous
experiments \cite{JPCnature}. It consists of five lines, two for each of the
idler and the signal modes serving as input and output, where both input and
output lines are connected to the sample through a cryogenic circulator at
base temperature. The fifth line carries the pump signal P. The input lines
include cryogenic attenuators at the $4$ $%
\operatorname{K}%
$ stage and base, while the output lines include base temperature circulators,
a cryogenic HEMT amplifier at the $4$ $%
\operatorname{K}%
$ stage and an additional amplification stage at room temperature. A small
magnetic coil positioned on top of the sample mount was used to flux bias the
JRM loop at the amplification working point $\varphi\simeq\varphi_{0}/2$,
where $\varphi_{0}=\hbar/2e$. Note that both P and I in the present design are
fed to the resonator of the J-amp via a $180$ degree hybrid as shown in Fig.
\ref{JPCdeviceN} (a), and not through direct capacitive coupling to the JRM as
in Ref. \cite{JPCnature}. This new method simplifies the design considerably
as it cancels the need of an on-chip wire cross-over.

The J-amp sample was fabricated on a $300%
\operatorname{\mu m}%
$ thick silicon chip having a resistivity larger than $1%
\operatorname{k\Omega }%
\times%
\operatorname{cm}%
$. An optical microscope image of the J-amp as well as SEM micrographs of the
JRM and one of the Josephson junctions are shown in Fig. \ref{JPCdeviceN} (b).
The microstrip resonators and the JRM are made of aluminum. The entire chip
was written using a one-layer e-beam lithography process. The Al/AlO$_{x}$/Al
Josephson junctions were evaporated using a standard shadow mask evaporation
process. The Josephson junction area is $5%
\operatorname{\mu m}%
\times1%
\operatorname{\mu m}%
$, while the loop area of the JRM is about $50%
\operatorname{\mu m}%
^{2}$. The critical current of the nominally identical Josephson junctions of
the JRM is $I_{0}=3\pm0.5%
\operatorname{\mu A}%
$. The spider-web-like structure of the top conductor of the microstrip
resonators was utilized in order to reduce potential losses due to vortex motion.%

\begin{figure}
[h]
\begin{center}
\includegraphics[
height=2.7596in,
width=4.2237in
]%
{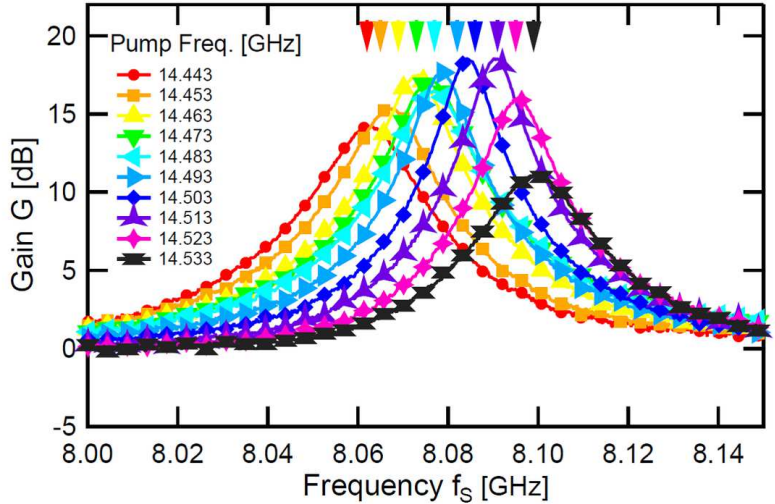}%
\caption{(Color online). Gain curves of the signal mode versus frequency. The
center frequency of the amplifier is tuned by varying the pump frequency. The
data show a frequency range of about $60$ $\operatorname{MHz}$ in which the
signal gain exceeds $10$ dB. The arrows plotted on top of the curves indicate
the locations of the calculated signal frequencies.}%
\label{SignalTunabilityN}%
\end{center}
\end{figure}

In Fig. \ref{SignalTunabilityN} we plot amplification curves of the signal as
a function of frequency measured in reflection. The gain curves shown in the
figure correspond to different pump frequencies as listed in the legend, where
for each applied pump frequency the pump power was adjusted to yield maximum
gain. The figure shows a maximum signal gain of about $18$ dB and a
center-frequency tunability of about $60$ $%
\operatorname{MHz}%
$ with gain in excess of $10$ dB in agreement with theory. Furthermore, due to
a good bandwidth matching between S and I resonators, similar frequency
tunability was achieved for the idler as well.%

\begin{figure}
[h]
\begin{center}
\includegraphics[
height=2.1404in,
width=4.2774in
]%
{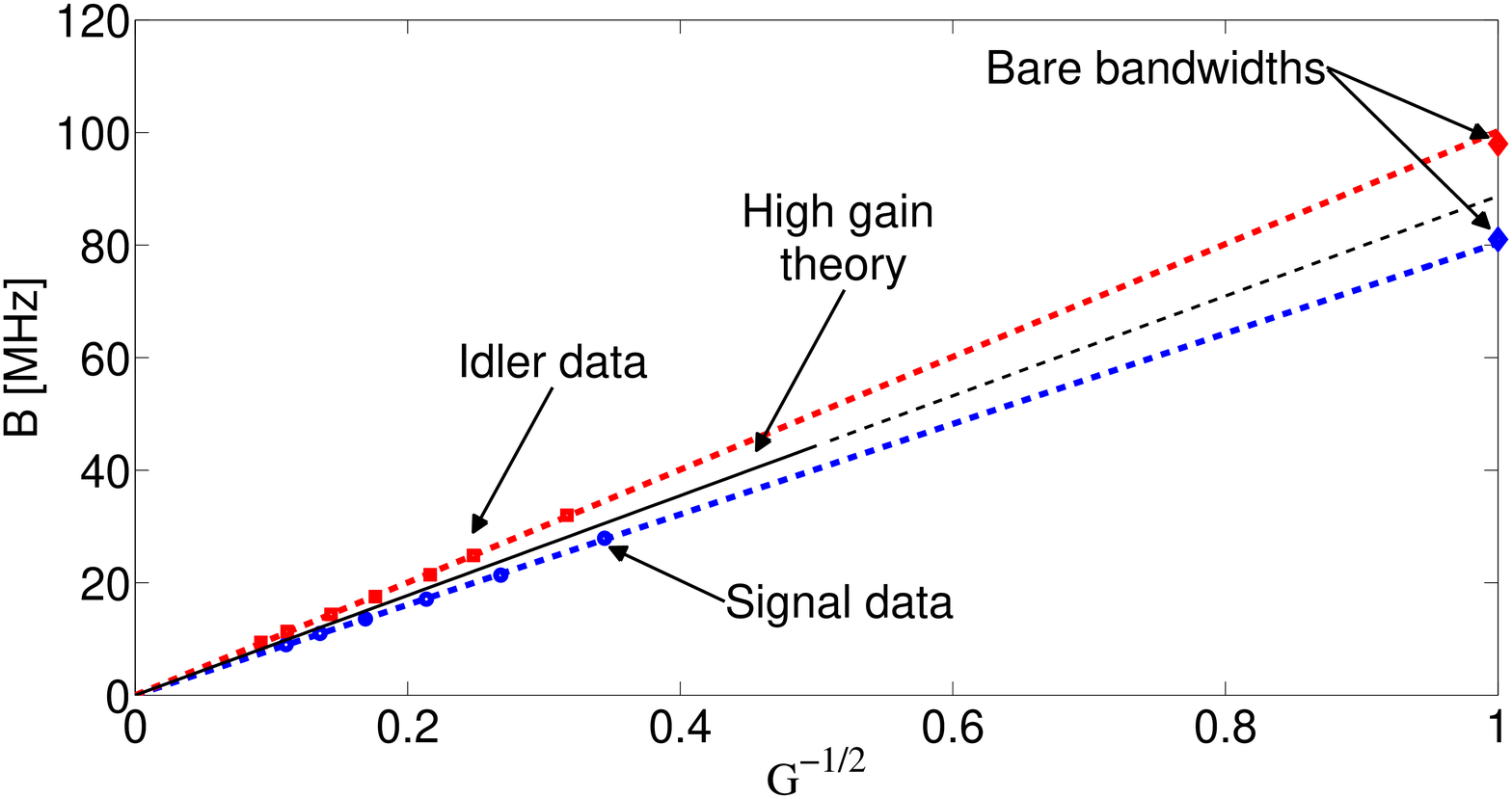}%
\caption{(Color online). The bandwidth of the amplifier $B$ versus $1/\sqrt
{G}$. The red (squares) and blue (circles) correspond to idler and signal data
respectively. The linear fits drawn as red and blue dashed lines satisfy a
relation of the form $B\sqrt{G}=const$. The black solid line depicts the
theoretical expression for the bandwidth $B$ of the amplifier in the high gain
limit (see text). The two diamond-shaped points red (top) and blue (bottom)
drawn at unity gain correspond to the bandwidths of the bare resonators I and
S respectively.}%
\label{bwvsgain}%
\end{center}
\end{figure}

The interplay between $G$ and $B$ of the device is shown in Fig.
\ref{bwvsgain}. According to J-amp theory \cite{JPCnaturePhys,JRCAreview}, in
the limit of large gains ($G\gg1$) the amplification bandwidth (corresponding
to the $-3$ dB points below the maximum) varies with the device gain as
$B=2(2\pi/\kappa_{I}+2\pi/\kappa_{S})^{-1}G^{-1/2}$. In Fig. \ref{bwvsgain}
the measured bandwidth of the J-amp at resonant tuning ($f_{P}=f_{S}%
^{res}+f_{I}^{res}$) is plotted versus $G^{-1/2}$ for both the signal (blue
circles) and the idler (red squares). The blue and red dashed lines are linear
fits to the data and they coincide to a good agreement with the theoretical
prediction for large gains (drawn as a solid black line) and also extrapolate
well to the bare bandwidths measured for the signal and idler resonators at
$\varphi=\varphi_{0}/2$ with no gain (pump off). At the maximum-gain points in
the figure (expressed in linear scale), the J-amp attains $19.1$ dB of gain
and $11.3$ $%
\operatorname{MHz}%
$ of bandwidth for the signal mode and $20.7$ dB and $9.4%
\operatorname{MHz}%
$ for the idler mode.

Moreover, by measuring the improvement in the signal to noise ratio of the
system due to the presence of the J-amp and the noise temperature of the
output lines, we are able to set an upper limit on the amount of noise added
by the J-amp to the input which was found to be $1\pm0.5$ photon at the signal frequency.

\begin{figure}
[h]
\begin{center}
\includegraphics[
height=2.3272in,
width=4.6726in
]%
{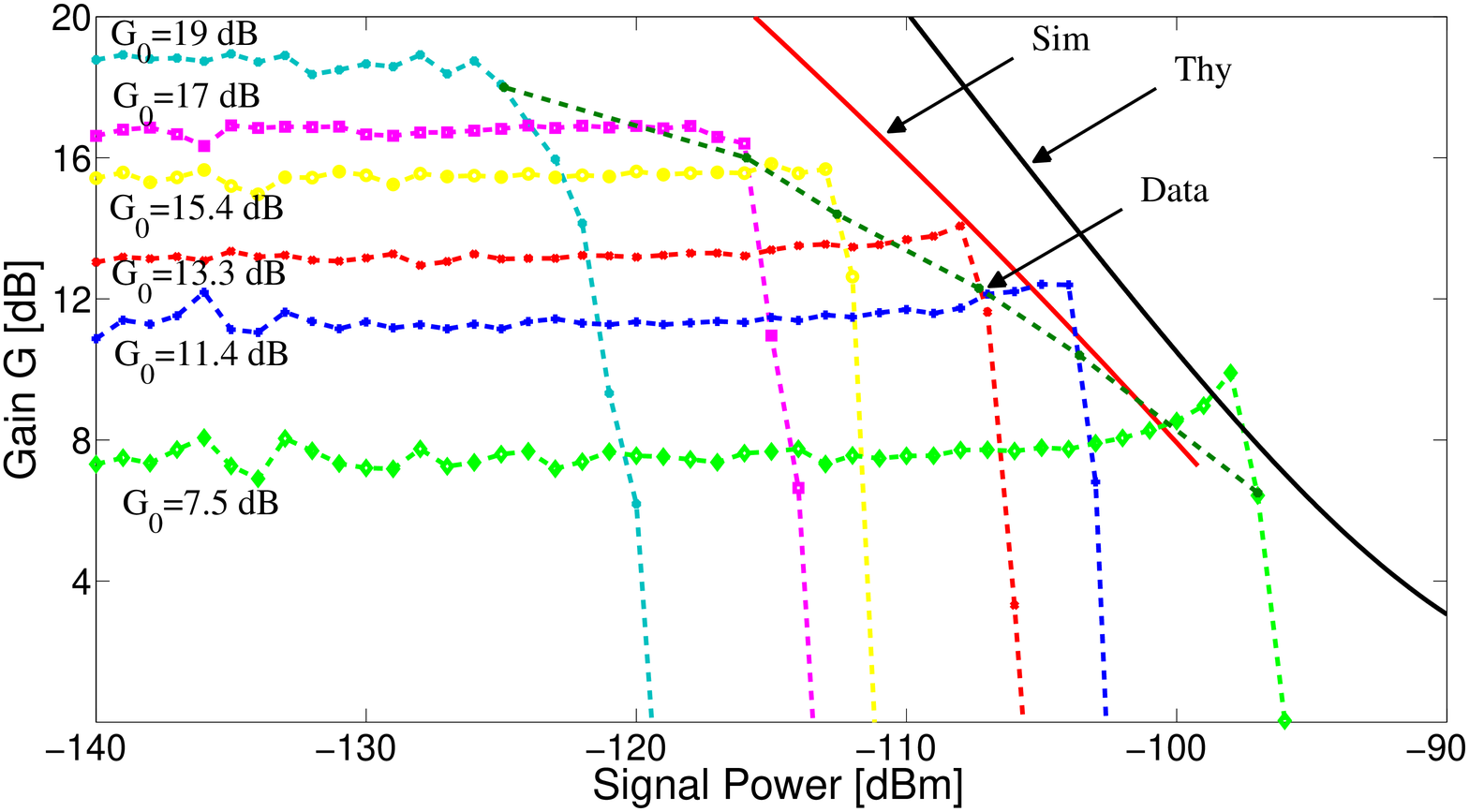}%
\caption{(Color online). A $1$ dB compression point measurement of the J-amp
as a function of signal gain. The dashed green line connects $1$ dB
compression points measured for different gains $G_{0}$. The black curve
depicts the upper limit on the signal power predicted by theory. The solid red
curve corresponds to a Microwave Office transmission-line simulation of the
device in the range of measured gains. }%
\label{DynamicRange1}%
\end{center}
\end{figure}

Another important figure of merit of the device is the $1$ dB compression
point, i.e. the maximum signal power at the input of the device, denoted as
$P_{\mathrm{\max}}$, at which the gain of the device drops by $1$ dB below
$G_{0}$ (the gain of the device for a vanishing input power). In Fig.
\ref{DynamicRange1} we plot the signal gain as a function signal power. The
different data sets correspond to different values of $G_{0}$. The $1$ dB
compression point of each data set is indicated in green. The dashed green
line connecting these points is a guide for the eye. The solid red curve
corresponds to a Microwave Office simulation of the J-amp based on a
transmission-line-model \cite{JPCnaturePhys,JRCAreview}.

The theoretical bound on the maximum signal power of the J-amp derived in Ref.
\cite{JRCAreview} $P_{\mathrm{\max}}=\left(  0.16\pi/p_{S}p_{I}Q_{S}%
Q_{I}\right)  \left(  \gamma_{P}\omega_{S}/\omega_{P}^{2}\right)  \left(
Z_{0}I_{0}^{2}/\left(  G_{0}-1\right)  \right)  $, is shown in Fig.
\ref{DynamicRange1} as a black curve, where $\omega_{S}$ and $\omega_{P}$ are
the angular frequencies of the signal and the pump, $p_{S,I}=L_{J}/\left(
L_{J}+L_{S,I}\right)  \simeq0.03-0.04$ is the participation ratio of the
device \cite{JPCnaturePhys,JRCAreview}, where $L_{J}$ is the effective
Josephson junction inductance at the working point given by $L_{J}=\sqrt
{2}\varphi_{0}/I_{0}$ and $L_{S,I}$ is the effective inductance of the S and I
resonators, $\gamma_{P}/2\pi$ $\simeq0.9%
\operatorname{GHz}%
$ is the residence rate of photons at $\omega_{P}$ and $Z_{0}=50%
\operatorname{\Omega }%
$ is the characteristic impedance of the feedline.

By comparing the dashed green line and the solid black or red curves, we see
that in the low gain limit both theory and simulation curves show a good
agreement with the data. However, in the high gain limit the slope of the
decrease in $P_{\mathrm{\max}}$ as a function of $G_{0}$ becomes larger than
$\sim-1$ dB/dB predicted by theory or $\sim$ $-1.3$ dB/dB given by simulation.
An important key to understanding this discrepancy is the behavior of the gain
in the vicinity of $P_{\mathrm{\max}}$. Steep drops in the gain similar to the
ones seen in Fig. \ref{DynamicRange1} suggest that the device enters a chaotic
regime at $P_{\mathrm{\max}}$ due to nonlinear processes taking place in the
Josephson junctions which are not accounted for in the simplified ideal
theoretical expression or in the simulation. Other devices, where the $pQ$
product was increased at the expense of the gain-bandwidth product, exhibited
a qualitatively different behavior near $P_{\mathrm{\max}}$. In these devices,
the drop in gain is gradual and the gain curves are in full agreement with
theory and simulation.

It is worthwhile noting that $P_{\mathrm{\max}}$ which can be processed by the
present device at $19$ dB of gain is $-125$ dBm, which corresponds to $5$
photons at the signal frequency times the amplifier dynamic bandwidth. Thus,
allowing for qubit state readout which is performed, in general, using one
photon on average in the readout cavity with a bandwidth of $20$ $%
\operatorname{MHz}%
$ \cite{fluxonium}.%

In order to enhance the $P_{\mathrm{\max}}$ figure of the device while
maintaining the current values for $G$, $\omega$ and $B$, $I_{0}$ should be
increased according to the theoretical expression. However, such increase in
$I_{0}$ implies decreasing $L_{S,I}$ and increasing the resonator and coupling
capacitance in order to keep $p$, $\omega$ and $Q$ constant. Consequently,
these constraints set an upper limit on $I_{0}$ which can be utilized with the
present design of microstrip resonators. Increasing $I_{0}$ by a factor of
$3.3$ ($I_{0}=10%
\operatorname{\mu A}%
$) for instance, would increase $P_{\mathrm{\max}}$ by $10$ dB, but would also
require decreasing the characteristic impedance to $15%
\operatorname{\Omega }%
$. In that case, it might be beneficial to replace the transmission-line
resonators with planar capacitors and semi-lumped inductors
\cite{JBA,Hatridge}.

The numerous practical properties of this ultra-low added noise amplifier
makes it suitable for the purpose of quantum nondemolition readout of
superconducting qubits \cite{FidelitySNR} which in turn leads to observation
of quantum jumps \cite{QuantumJumps} and quantum feedback.

Discussions with R. J. Schoelkopf, L. Frunzio, N. Bergeal and B. Huard are
gratefully acknowledged. This research was supported by the NSF under grants
DMR-0653377; the NSA through ARO Grant No. W911NF-09-1-0514, IARPA under ARO
Contract No. W911NF-09-1-0369, the Keck foundation, and Agence Nationale pour
la Recherche under grant ANR07-CEXC-003. M.H.D. acknowledges partial support
from College de France.






\bibliographystyle{plain}
\bibliography{book}

\end{document}